\documentstyle[11pt]{article}
\begin{document}

\title{\textbf{The generalized second law of gravitational thermodynamics on the apparent and event
horizons in FRW cosmology}}
\author{K. Karami$^{1,2}$\thanks{E-mail: KKarami@uok.ac.ir} ,
S. Ghaffari${^1}$ , M.M. Soltanzadeh${^1}$\\$^{1}$\small{Department
of Physics, University of Kurdistan, Pasdaran St., Sanandaj,
Iran}\\$^{2}$\small{Research Institute for Astronomy
$\&$ Astrophysics of Maragha (RIAAM), Maragha, Iran}\\
}

\maketitle

\begin{abstract}
We investigate the validity of the generalized second law of
gravitational thermodynamics on the apparent and event horizons in a
non-flat FRW universe containing the interacting dark energy with
dark matter. We show that for the dynamical apparent horizon, the
generalized second law is always satisfied throughout the history of
the universe for any spatial curvature and it is independent of the
equation of state parameter of the interacting dark energy model.
Whereas for the cosmological event horizon, the validity of the
generalized second law depends on the equation of state parameter of
the model.
\end{abstract}
\clearpage
\section{Introduction}
The present acceleration of the universe expansion has been well
established through numerous and complementary cosmological
observations \cite{Riess}. A component which is responsible for this
accelerated expansion usually dubbed ``dark energy'' (DE). However,
the nature of DE is still unknown, and people have proposed some
candidates to describe it (for review see
\cite{Padmanabhan,Copeland} and references therein).

One of the important questions in cosmology concerns the
thermodynamical behavior of the accelerated expanding universe
driven by DE. It was shown that the Einstein equation can be derived
from the first law of thermodynamics by assuming the proportionality
of entropy and the horizon area \cite{Hawking,Bardeen,Jacobson}. In
the cosmological context, attempts to disclose the connection
between Einstein gravity and thermodynamics were carried out. It was
shown that the differential form of the Friedmann equation in the
Friedmann-Robertson-Walker (FRW) universe can be written in the form
of the first law of thermodynamics on the apparent horizon
\cite{Cai05}. Further studies on the equivalence between the first
law of thermodynamics and Friedmann equation has been investigated
in various gravity theories like Gauss-Bonnet, Lovelock and
braneworld scenarios \cite{Cai05,Akbar,Sheykhi2}.

Besides examining the validity of the thermodynamical interpretation
of gravity by expressing the gravitational field equations into the
first law of thermodynamics in different spacetimes, it is also of
great interest to investigate the validity of the generalized second
law (GSL) of thermodynamics in the accelerating universe driven by
DE. The GSL of thermodynamics is as important as the first law,
governing the development of the nature
\cite{Davies,Gong,Izquierdo2,Izquierdo,Zhau,Sheykhi3,Jamil,Wang2,Karami2,Karami3}.

Here our aim is to investigate the validity of the GSL of
gravitational thermodynamics for the interacting DE model with DM in
a non-flat FRW universe enclosed by the dynamical apparent horizon
and the cosmological event horizon. Note that in the literature,
people usually have studied the validity of the GSL for a specific
model of DE with special kind of interaction term. But we would like
to extend it to any DE model with general interaction term. This
paper is organized as follows. In section \ref{II}, we study the DE
model in a non-flat FRW universe which is in interaction with the
DM. In section \ref{III}, we investigate the validity of the GSL of
gravitational thermodynamics for the universe enclosed by the
apparent horizon and the event horizon which is in thermal
equilibrium with the Hawking radiation of the horizon. Also we give
an example for the case of the cosmological event horizon. In
section \ref{IV} we investigate the effect of interaction term
between DE and DM on the GSL. Section \ref{V} is devoted to
conclusions.
\section{Interacting DE and DM in FRW cosmology}\label{II}

The first Friedmann equation in FRW cosmology takes the form
\begin{equation}
{\textsl{H}}^2+\frac{k}{a^2}=\frac{8\pi}{3}~
(\rho_{\Lambda}+\rho_{\rm m}),\label{eqfr}
\end{equation}
where we take $G=1$ and $k=0,1,-1$ represent a flat, closed and open
FRW universe, respectively. Also $\rho_{\Lambda}$ and $\rho_{\rm m}$
are the energy density of DE and DM, respectively.

From (\ref{eqfr}), we can write
\begin{equation}
\Omega_{\rm m}+\Omega_{\Lambda}=1+\Omega_{k},\label{eq10}
\end{equation}
where we have used the following definitions
\begin{equation}
\Omega_{\rm m}=\frac{8\pi\rho_{\rm m}}{3H^2},~~\Omega_{\rm
\Lambda}=\frac{8\pi\rho_{\Lambda}}{3H^2},~~\Omega_{k}=\frac{k}{a^2H^2}.
\label{eqomega}
\end{equation}
Since we consider the interaction between DE and DM,
$\rho_{\Lambda}$ and $\rho_{\rm m}$ do not conserve separately.
Hence the energy conservation equations for DE and DM are
\begin{equation}
\dot{\rho}_{\Lambda}+3H(1+\omega_{\Lambda})\rho_{\Lambda}=-Q,\label{eqpol}
\end{equation}
\begin{equation}
\dot{\rho}_{\rm m}+3H\rho_{\rm m}=Q,\label{eqCDM}
\end{equation}
where Q stands for the interaction term. For $Q>0$, there is an
energy transfer from the DE to DM. The choice of the interaction
between both components was to get a scaling solution to the
coincidence problem such that the universe approaches a stationary
stage in which the ratio of DE and DM becomes a constant
\cite{Huang}. The dynamics of interacting DE models with different
$Q$-classes have been studied in ample detail by \cite{Amendola}.
Here we continue our study without considering a specific form for
the interaction term.

Taking the time derivative in both sides of Eq. (\ref{eqfr}), and
using Eqs. (\ref{eq10}), (\ref{eqomega}), (\ref{eqpol}) and
(\ref{eqCDM}), one can get the equation of state (EoS) parameter of
DE as
\begin{equation}
\omega_{\Lambda}=-\frac{1}{3\Omega_{\Lambda}}\Big(3+\Omega_{k}+\frac{2\dot{H}}{H^2}\Big).\label{wlambda}
\end{equation}
For the DE dominated universe, i.e. $\Omega_{\Lambda}\rightarrow 1$,
the above equation yields
\begin{equation}
\omega_{\Lambda}=-1-\frac{2\dot{H}}{3H^2},
\end{equation}
which shows that for the phantom, $\omega_{\Lambda}<-1$, and
quintessence, $\omega_{\Lambda}>-1$, dominated universe, we need to
have $\dot{H}>0$ and $\dot{H}<0$, respectively. Note that for the de
Sitter universe, i.e. $\dot{H}=0$, we have $\omega_{\Lambda}=-1$
which behaves like the cosmological constant.

The deceleration parameter is given by
\begin{equation}
q=-\Big(1+\frac{\dot{H}}{H^2}\Big).\label{q1}
\end{equation}
Replacing the term $\dot{H}/H^2$ from (\ref{wlambda}) into
(\ref{q1}) yields
\begin{equation}
q=\frac{1}{2}\Big(1+\Omega_{k}+3\Omega_{\Lambda}\omega_{
\Lambda}\Big).\label{q2}
\end{equation}



\section{GSL with corrected entropy-area relation}\label{III}

Here, we study the validity of the GSL of gravitational
thermodynamics. According to the GSL, entropy of matter and fluids
inside the horizon plus the entropy of the horizon do not decrease
with time \cite{Wang2}. Here like \cite{Sheykhi3}, we assume that
the temperature of both dark components are equal, due to their
mutual interaction. Note that Lima and Alcaniz \cite{Lima} using a
very naive estimate obtained the present value of the DE temperature
as $T_{\rm DE}^0\sim 10^{-6}~{\rm K}$. Also Zhou et al. \cite{Zhou}
estimated the DM temperature as $T_{\rm DM}^0\sim 10^{-7}~\rm K$ for
the present time. Since the temperature of the DE at the present
time differs from that of the DM, the systems must interact for some
length of time before they can attain thermal equilibrium. Although
in this case the local equilibrium hypothesis may no longer hold
\cite{Zhou,Das}, Karami and Ghaffari \cite{Karami1} showed that the
contribution of the heat flow between the DE and the DM in the GSL
in non-equilibrium thermodynamics is very small, $O(10^{-7})$.
Therefore the equilibrium thermodynamics is still preserved. We also
limit ourselves to the assumption that the thermal system including
the DE and DM bounded by the horizon remain in equilibrium so that
the temperature of the system must be uniform and the same as the
temperature of its boundary. This requires that the temperature $T$
of the both DE and DM inside the horizon should be in equilibrium
with the Hawking temperature $T_{\rm h}$ associated with the
horizon, so we have $T = T_{\rm h}$. This expression holds in the
local equilibrium hypothesis. If the temperature of the system
 differs much from that of the horizon, there will be spontaneous heat flow between
  the horizon and the fluid and the local equilibrium hypothesis will no
longer hold \cite{Zhou,Das,Karami1}. This is also at variance with
the FRW geometry. In general, when we consider the thermal
equilibrium state of the universe, the temperature of the universe
is associated with the horizon.

The entropy of the universe including the DE and DM inside the
horizon can be related to its energy and pressure in the horizon by
Gibb's equation \cite{Izquierdo}
\begin{equation}
T{\rm d}S={\rm d}E+P{\rm d}V,\label{eqSLT1}
\end{equation}
where ${\rm V}=4\pi R_{\rm h}^3/3$ is the volume containing the DE
and DM with the radius of the horizon $R_{\rm h}$ and $T=T_{\rm
h}=1/(2\pi R_{\rm h})$ is the Hawking temperature of the horizon.
Also
\begin{equation}
E=\frac{4\pi R_{\rm h}^3}{3} (\rho_{\Lambda}+\rho_{\rm
m}),\label{eqEde}
\end{equation}
\begin{equation}
P=P_{\Lambda}+P_{\rm
m}=P_{\Lambda}=\omega_{\Lambda}\rho_{\Lambda}=\frac{3H^2}{8\pi}\omega_{\Lambda}\Omega_{\Lambda}.
\label{eqEcdm}
\end{equation}
Taking the derivative in both sides of (\ref{eqSLT1}) with respect
to cosmic time $t$, and using Eqs. (\ref{eqfr}), (\ref{eq10}),
(\ref{eqomega}), (\ref{eqpol}), (\ref{eqCDM}), (\ref{eqEde}) and
(\ref{eqEcdm}), we obtain the evolution of the entropy in the
universe containing the DE and DM as
\begin{equation}
\dot{S}=3\pi H^{2}R_{\rm h}^2(R_{\rm h}\dot{R_{\rm h}}-HR_{\rm
h}^{2})(1+\Omega_k+\Omega_{\Lambda}\omega_{\Lambda}).\label{S2}
\end{equation}
Also in addition to the entropy in the universe, there is a
geometric entropy on the horizon $S_{{\rm h}}=\pi R_{\rm h}^{2}$
\cite{Izquierdo}. The evolution of this horizon entropy is obtained
as
\begin{equation}
\dot{S}_{\rm h}=2\pi R_{\rm h}\dot{R_{\rm h}}.\label{Sh}
\end{equation}
Finally, the GSL due to different contributions of the DE, DM and
horizon is obtained as
\begin{equation}
\dot{S}_{\rm tot}=3\pi H^{2}R_{\rm h}^{2}(R_{\rm h}\dot{R_{\rm
h}}-HR_{\rm
h}^{2})(1+\Omega_k+\Omega_{\Lambda}\omega_{\Lambda})+2\pi R_{\rm
h}\dot{R_{\rm h}},\label{Stot2}
\end{equation}
where $S_{\rm tot}=S+S_{\rm h}$ is the total entropy. Note that in
$S_{\rm tot}$ we ignored the contribution of baryonic matter (BM)
($\Omega^0_{BM}\sim 0.04$) in comparison with the DM and DE
($\Omega^0_{DM}+\Omega^0_{DE}\sim 0.96$). Because according to the
recent measurements of the supermassive black hole mass function,
the present entropy of BM $S_{BM}^0=(2.7\pm 2.1)\times 10^{80}$ is
five to seven orders of magnitude smaller than the DM
$S_{DM}^0=6\times 10^{86\pm 1}$ \cite{Egan}.

In the next sections, we investigate the validity of the GSL given
by Eq. (\ref{Stot2}) for the dynamical apparent and cosmological
event horizons.
\subsection{The dynamical apparent horizon}

The dynamical apparent horizon in the FRW universe is given by
\cite{Cai09}
\begin{equation}
R_{\rm A}=H^{-1}(1+\Omega_{k})^{-1/2}.\label{ah}
\end{equation}
For $k = 0$, the apparent horizon is same as the Hubble horizon.

Recently Cai et al. \cite{Cai09} proofed that the apparent horizon
of the FRW universe with any spatial curvature has indeed an
associated Hawking temperature $T_{\rm A} = 1/2\pi R_{\rm A}$. Cai
et al. \cite{Cai09} also showed that the Hawking temperature can be
measured by an observer with the Kodoma vector inside the apparent
horizon.

If we take the derivative in both sides of (\ref{ah}) with respect
to cosmic time $t$, then we obtain
\begin{equation}
\dot{R_{\rm
A}}=\frac{3(1+\Omega_{k}+\Omega_{\Lambda}\omega_{\Lambda})}{2(1+\Omega_{k})^{3/2}}\label{ahdot}.
\end{equation}
Using Eqs. (\ref{ah}) and (\ref{ahdot}) one can get
\begin{equation}
R_{\rm A}\dot{R_{\rm A}}-HR_{\rm
A}^{2}=\frac{(1+\Omega_k+3\Omega_{\Lambda}\omega_{\Lambda})}{2H(1+\Omega_{k})^2}\label{ahahdot}.
\end{equation}
Substituting Eqs. (\ref{ah}), (\ref{ahdot}) and (\ref{ahahdot}) in
(\ref{S2}) and (\ref{Sh}) reduce to
\begin{equation}
\dot{S}=\frac{3\pi}{2H(1+\Omega_{k})^3}(1+\Omega_{k}+3\Omega_{\Lambda}\omega_{\Lambda})
(1+\Omega_{k}+\Omega_{\Lambda}\omega_{\Lambda}),\label{Sah}
\end{equation}
\begin{eqnarray}
\dot{S}_{\rm
A}=\frac{3\pi}{H(1+\Omega_{k})^2}(1+\Omega_{k}+\Omega_{\Lambda}\omega_{\Lambda}).\label{SAah}
\end{eqnarray}
Equation (\ref{Sah}) shows that for
$-(\frac{1+\Omega_{k}}{\Omega_{\Lambda}})<\omega_{\Lambda}<-\frac{1}{3}(\frac{1+\Omega_{k}}{\Omega_{\Lambda}})$,
the contribution of entropy of the universe inside the dynamical
apparent horizon in the GSL is negative, i.e. $\dot{S}<0$. For the
late-time or the DE dominated universe where
$\Omega_{\Lambda}\rightarrow 1$ and $R_{A}=H^{-1}$, the entropy of
the universe will be a non-increasing function of time in the
quintessence regime with $-1<\omega_{\Lambda}<-1/3$.

Equation (\ref{SAah}) clears that for
$\omega_{\Lambda}>-(\frac{1+\Omega_{k}}{\Omega_{\Lambda}})$, the
contribution of the dynamical apparent horizon in the GSL is
positive, i.e. $\dot{S}_{\rm A}>0$. For the DE dominated universe,
the entropy of the dynamical apparent horizon will be an increasing
function of time in the quintessence regime with
$\omega_{\Lambda}>-1$.

Finally, using Eqs. (\ref{Sah}) and (\ref{SAah}), the GSL due to
different contributions of the DE, DM and apparent horizon can be
obtained as
\begin{equation}
\dot{S}_{\rm
tot}=\frac{9\pi}{2H(1+\Omega_{k})^3}(1+\Omega_{k}+\Omega_{\Lambda}\omega_{\Lambda})^2\geq
0.\label{Stotah}
\end{equation}
Equation (\ref{Stotah}) presents that the GSL for the universe
containing the interacting DE with DM enclosed by the dynamical
apparent horizon is always satisfied throughout the history of the
universe for any spatial curvature and it is independent of the EoS
parameter of the interacting DE model.
\subsection{The cosmological event horizon}

For the cosmological event horizon defined as
\begin{equation}
R_{\rm E}=a\int_{t}^{\infty}\frac{{\rm
d}t}{a}=a\int_{a}^{\infty}\frac{{\rm d}a}{Ha^2},\label{eh}
\end{equation}
one can obtain
\begin{equation}
\dot{R}_{\rm E}=HR_{\rm E}-1.\label{ehdot}
\end{equation}
For a de Sitter space-time where $H$ is a constant, Eqs. (\ref{eh})
and (\ref{ehdot}) show that the cosmological event horizon radius is
$H^{-1}$ and $\dot{R}_{\rm E}=0$. Therefore, in a spatially flat de
Sitter universe, the event horizon and the apparent horizon, given
by Eq. (\ref{ah}), of the universe coincide with each other and
there is only one cosmological horizon \cite{Gong}. For the de
Sitter universe, from Eq. (\ref{wlambda}) we have
$\omega_{\Lambda}=-1/\Omega_{\Lambda}$ hence Eq. (\ref{Stot2}) shows
that $\dot{S}_{\rm tot}=\dot{S}=0$, which corresponds to a
reversible adiabatic expansion.

Substituting Eq. (\ref{ehdot}) into (\ref{S2}) yields the entropy of
the universe inside the cosmological event horizon as
\begin{equation}
\dot{S}=-3\pi H^{2}R_{\rm
E}^3(1+\Omega_k+\Omega_{\Lambda}\omega_{\Lambda}).\label{Seh}
\end{equation}
Equation (\ref{Seh}) clears that for
$\omega_{\Lambda}<-(\frac{1+\Omega_{k}}{\Omega_{\Lambda}})$, the
contribution of the cosmological event horizon in the GSL is
positive, i.e. $\dot{S}>0$. For the late-time universe, the entropy
of the universe will be an increasing function of time in the
phantom regime with $\omega_{\Lambda}<-1$.

Following \cite{Sadjadi}, for the quintessence and phantom universe,
$\dot{R}_{\rm E}>0$ and $\dot{R}_{\rm E}<0$, respectively.
Therefore, form Eq. (\ref{Sh}) one can conclude that for the
quintessence universe $\dot{S}_{\rm E}=2\pi R_{\rm E}\dot{R_{\rm
E}}>0$ and for the phantom universe $\dot{S}_{\rm E}<0$.

Substituting Eq. (\ref{ehdot}) into (\ref{Stot2}) yields
\begin{equation}
\dot{S}_{\rm tot}=2\pi R_{\rm E}\left[\dot{R}_{\rm
E}-\frac{3}{2}(1+\Omega_k+\Omega_{\Lambda}\omega_{\Lambda})H^2R^2_{E}\right],\label{Stoteh}
\end{equation}
which shows that the GSL is satisfied, i.e. $\dot{S}_{\rm tot}\geq
0$, when
\begin{equation}
\omega_{\Lambda}\leq\frac{2\dot{R}_{\rm
E}}{3H^2R^2_{E}\Omega_{\Lambda}}-\Big(\frac{1+\Omega_{k}}{\Omega_{\Lambda}}\Big).\label{Stoteh2}
\end{equation}
The above constraint on the EoS parameter of the interacting DE
model presents that for the late-time universe in the quintessence
and phantom regime, we have $\omega_{\Lambda}\leq\frac{2\dot{R}_{\rm
E}}{3H^2R^2_{E}}-1$ and $\omega_{\Lambda}\leq-\frac{2|\dot{R}_{\rm
E}|}{3H^2R^2_{E}}-1$, respectively.

\begin{figure}
\includegraphics{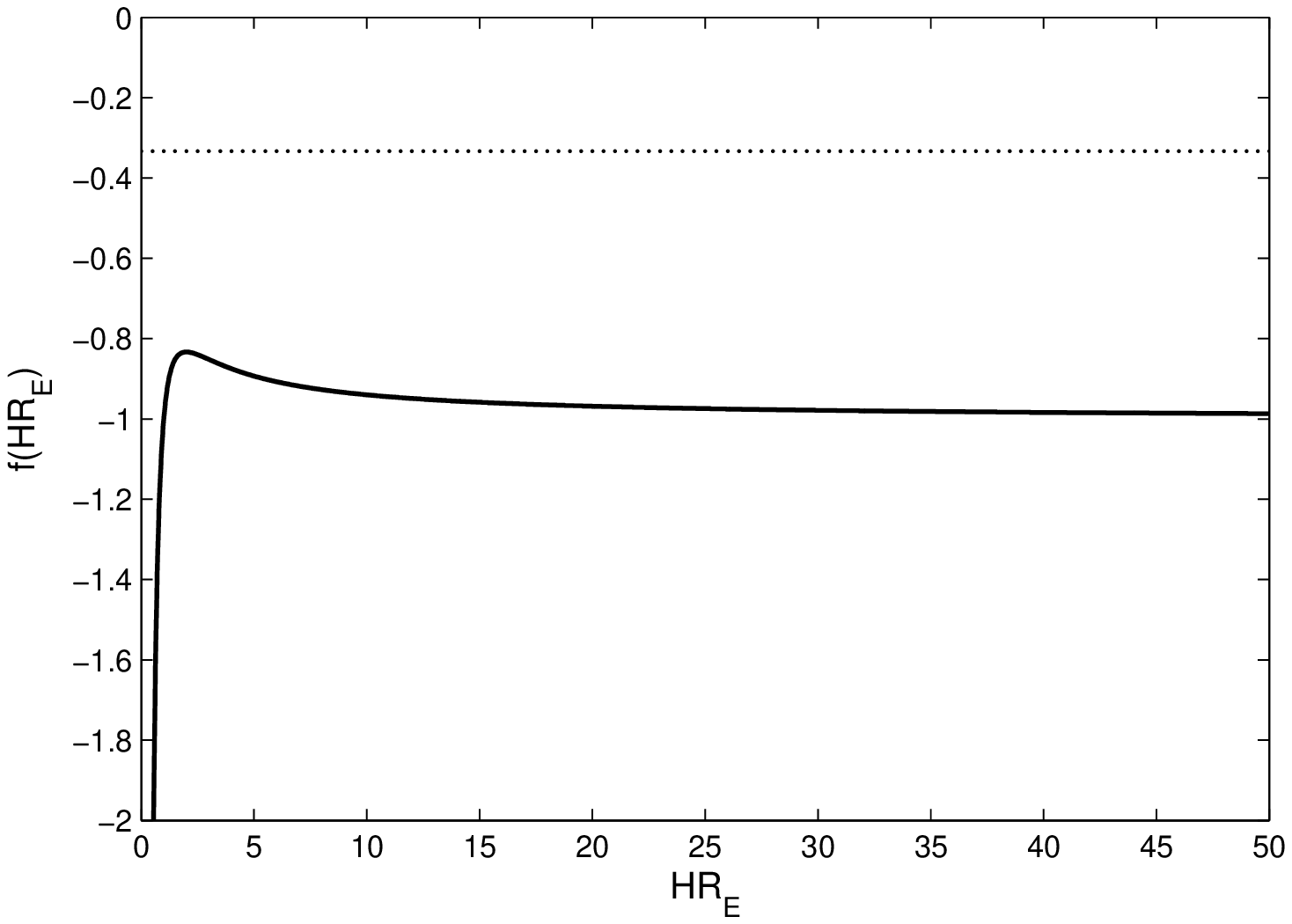}
      \vspace{6.5cm}
      \caption[]{$f(HR_E)=\frac{2(HR_E-1)}{3H^2R^2_{E}}-1$
      versus $HR_E$ for the DE dominated universe. The dotted line shows $f(HR_E)=-1/3$.}
         \label{f}
   \end{figure}

Substituting Eq. (\ref{ehdot}) into (\ref{Stoteh2}) yields
$\omega_{\Lambda}\leq f(HR_E)$ where
\begin{equation}
f(HR_E)=\frac{2(HR_E-1)}{3H^2R^2_{E}\Omega_{\Lambda}}-\Big(\frac{1+\Omega_{k}}{\Omega_{\Lambda}}\Big).\label{Stoteh3}
\end{equation}
In Fig. \ref{f}, the solid line shows $f(HR_E)$ versus $HR_E$ for
the DE dominated universe where $\Omega_\Lambda\rightarrow 1$ and
the dotted line presents $f(HR_E)=-1/3$. Note that for a universe
enclosed by the event horizon, we have always
$\omega_{\Lambda}<-1/3$. Figure \ref{f} clears that for
$f(HR_E)<\omega_{\Lambda}<-1/3$ the GSL is not satisfied on the
cosmological event horizon and it remains valid only for
$\omega_{\Lambda}\leq f(HR_E)$.

Therefore, for the non-flat FRW universe containing the interacting
DE with DM enclosed by the cosmological event horizon, the GSL is
satisfied for the special range of the EoS parameter of the DE
model. Contrary to the case of the apparent horizon, the validity of
the GSL for the cosmological event horizon depends on the EoS
parameter of the interacting DE model.
\subsection{A pole-like type phantom universe enclosed by the event
horizon}

Here we give an example to study the GSL in the case of the
cosmological event horizon. Following \cite{Sadjadi}, we consider a
phantom DE model of the universe describing by a pole-like type
scale factor as
\begin{equation}
a(t)=a_0(t_s-t)^{-n},~~~~~~t\leq t_s,~~~~~~n>0,\label{aexa}
\end{equation}
then one can get
\begin{equation}
H=\frac{n}{t_s-t},\label{H}
\end{equation}
also
\begin{equation}
\dot{H}=\frac{n}{(t_s-t)^2}>0.\label{hdot}
\end{equation}
The cosmological event horizon can be obtained as
\begin{equation}
R_{\rm E}=a\int_{t}^{t_s}\frac{{\rm
d}t}{a}=\frac{t_s-t}{n+1},\label{reexa}
\end{equation}
also
\begin{equation}
\dot{R}_{\rm E}=-\frac{1}{n+1}<0.\label{redotexa}
\end{equation}
Equations (\ref{hdot}) and (\ref{redotexa}) confirm that the model
(\ref{aexa}) is correspondence to a phantom dominated universe.

The EoS and deceleration parameters of the model (\ref{aexa}) are
obtained by the help of Eqs. (\ref{wlambda}) and (\ref{q1}),
respectively, as
\begin{equation}
\omega_{\Lambda}=-\frac{1}{3\Omega_{\Lambda}}\Big(3+\Omega_{k}+\frac{2}{n}\Big),\label{wlambdaexa}
\end{equation}
\begin{equation}
q=-1-\frac{1}{n}<-1.
\end{equation}

Equation (\ref{wlambdaexa}) shows that for the late-time universe,
 we have
$\omega_{\Lambda}=-1-\frac{2}{3n}<-1$ which is the EoS parameter of
phantom DE.

From Eqs. (\ref{Sh}), (\ref{Seh}), (\ref{H}), (\ref{reexa}),
(\ref{redotexa}) and (\ref{wlambdaexa}), the entropy of the event
horizon and the entropy of the universe inside the event horizon can
be obtained as
\begin{equation}
\dot{S}_{\rm E}=2\pi R_{\rm E}\dot{R}_{\rm
E}=-\frac{2\pi}{(n+1)^2}(t_s-t)\leq 0,\label{Seexa}
\end{equation}
\begin{equation}
\dot{S}=\frac{2\pi
n^2}{(n+1)^3}(t_s-t)\Big(\frac{1}{n}-\Omega_k\Big).\label{Sexa}
\end{equation}
Equation (\ref{Seexa}) shows that the entropy of the event horizon
for the model (\ref{aexa}) has a negative contribution in the GSL
throughout the history of the universe. Equation (\ref{Sexa}) clears
that the entropy of the universe has a positive contribution in the
GSL only when $\Omega_{k}\leq 1/n$. Finally for the model
(\ref{aexa}), the GSL yields
\begin{equation}
\dot{S}_{\rm tot}=\dot{S}+\dot{S}_{\rm E}=-\frac{2\pi
(t_s-t)}{(n+1)^3}(1+n^2\Omega_k)<0,
\end{equation}
which clears that for the positive spatial curvature, compatible
with the present observations \cite{Bennett}, the GSL breaks down.

\section{The effect of interaction term between DE and DM on the GSL}\label{IV}
In our previous analysis, the interaction term between DE and DM is
not appeared explicitly in the GSL. To see how the DE-DM interaction
influences the GSL, we need to incorporate a specific form of the DE
model in our analysis. To do this we consider the holographic DE
(HDE) model which is motivated from the holographic principle
\cite{Hooft}. Following \cite{Huang}, the HDE density in a closed
universe is given by
\begin{equation}
\rho_{\Lambda}=3c^2M^2_PL^{-2},\label{hol}
\end{equation}
where $c$ is a positive constant and $M_P$ is the reduced Planck
Mass $M_P^{-2}=8\pi$. Recent observational data, which have been
used to constrain the HDE model, show that for the non-flat universe
$c=0.815_{-0.139}^{+0.179}$ \cite{Li5}. Also $L$ is the IR cut-off
defined as
\begin{equation}
L=\frac{a}{\sqrt{k}}\sin y,\label{L}
\end{equation}
where $y=\sqrt{k}R_{\rm E}/a$.  Note that $R_{\rm E}$ is the radial
size of the event horizon measured in the $r$ direction and $L$ is
the radius of the event horizon measured on the sphere of the
horizon \cite{Huang}. For the flat universe $L=R_{\rm E}$. For a
specific form of the interaction term between DE and DM as
$Q=3b^2H(\rho_{\Lambda}+\rho_{\rm m})$ with $b^2$ the coupling
constant \cite{Kim06}, the EoS parameter for the interacting HDE
with DM in a non-flat FRW universe is obtained as \cite{Lin}
\begin{equation}
\omega_{\Lambda}=-\frac{1}{3}-\frac{2\sqrt{\Omega_{\Lambda}}}{3c}\cos
y-b^2\Big(\frac{1+\Omega_{k}}{\Omega_{\Lambda}}\Big),\label{omegaeffh}
\end{equation}
where $\cos y=\sqrt{1-c^2\Omega_{k}/\Omega_{\Lambda}}$. Therefore
the coupling constant $b^2$ due to interaction is appeared
explicitly in the EoS parameter of the HDE. For the dynamical
apparent horizon, Eq. (\ref{Stotah}) shows that the GSL is always
satisfied throughout the history of the universere regardless of the
specific form of DE model and interaction term $Q$. But for the
cosmological event horizon, the story is different. The GSL for the
interacting HDE with DM in a non-flat universe enclosed by the event
horizon measured from the sphere of the horizon $L$ is obtained from
Eq. (\ref{Stot2}) as
\begin{equation}
\dot{S}_{\rm tot}=3\pi
H^{2}L^{2}(L\dot{L}-HL^{2})(1+\Omega_k+\Omega_{\Lambda}\omega_{\Lambda})+2\pi
L\dot{L},\label{Stot2L}
\end{equation}
where the necessary expressions for $L$ and $\dot{L}$ are given by
Eqs. (15) and (16) in \cite{Huang}, respectively, as
\begin{equation}
L=\frac{c}{H\sqrt{\Omega_{\Lambda}}},\label{L}
\end{equation}
\begin{equation}
\dot{L}=\frac{c}{\sqrt{\Omega_{\Lambda}}}-\cos{y}.\label{Ldot}
\end{equation}
Substituting Eqs. (\ref{omegaeffh}), (\ref{L}) and (\ref{Ldot}) in
(\ref{Stot2L}) yields
\begin{equation}
\dot{S}_{\rm tot}=\frac{\pi
c^3}{H\Omega_{\Lambda}^{3/2}}\left\{2\frac{\sqrt{\Omega_{\Lambda}}}{c}(1+\Omega_{\Lambda}\cos^2{y})+\Big(1-\frac{2}{c^2}\Big)\Omega_{\Lambda}\cos{y}
+3(b^2-1)(1+\Omega_{k})\cos{y} \right\},\label{dStot3L}
\end{equation}
which is same as the result given by Eq. (1.6) in \cite{Karami2}.
Equation (\ref{dStot3L}) shows that the coupling constant $b^2$ of
the interaction term $Q$ does affect the GSL on the radius of the
event horizon $L$. For instance, for $\cos y=0.99$,
$\Omega_{\Lambda}=0.73$, $\Omega_{k}=0.01$ and $c=1$ given by
\cite{Setare} for the present time, we get
\begin{equation}
\dot{S}_{\rm tot}=\frac{4.809\pi}{H}(b^2-0.264),
\end{equation}
which shows that only for $b^2\geq 0.264$ then $\dot{S}_{\rm
tot}\geq 0$ and the GSL is satisfied.

\section{Conclusions}\label{V}
Here the GSL of gravitational thermodynamics for the interacting DE
with DM in a non-flat FRW universe is investigated. Some
experimental data have implied that our universe is not a perfectly
flat universe and it possess a small positive curvature
\cite{Bennett}. Although it is believed that our universe is flat, a
contribution to the Friedmann equation from spatial curvature is
still possible if the number of e-foldings is not very large
\cite{Huang}. The boundary of the universe is assumed to be
enveloped by the dynamical apparent horizon and the cosmological
event horizon. We assumed that the universe to be in thermal
equilibrium with the Hawking temperature on the horizon. We found
that for the dynamical apparent horizon, the GSL is respected
throughout the history of the universe for any spatial curvature and
it is independent of the EoS parameter of the interacting DE model.
But for the cosmological event horizon, the GSL is satisfied for the
special range of the EoS parameter of the model.

The above results show that the dynamical apparent horizon in
comparison with the cosmological event horizon, is a good boundary
for studying cosmology, since on the apparent horizon there is the
well known correspondence between the first law of thermodynamics
and the Einstein equation \cite{Gong5}. In the other words, the
Friedmann equations describe local properties of spacetimes and the
apparent horizon is determined locally, while the cosmological event
horizon, Eq. (\ref{eh}), is determined by global properties of
spacetimes \cite{Cai05}. Besides in the dynamic spacetime, the
horizon thermodynamics is not as simple as that of the static
spacetime. The event horizon and apparent horizon are in general
different surfaces. The definition of thermodynamical quantities on
the cosmological event horizon in the nonstatic universe are
probably ill-defined \cite{Wang2}.
\\
\\
\noindent{{\bf Acknowledgements}}\\ The authors thank the reviewers
for very valuable comments. The work of K. Karami has been supported
financially by Research Institute for Astronomy $\&$ Astrophysics of
Maragha (RIAAM), Maragha, Iran.

\clearpage

\end{document}